\documentclass[twocolumn,floatfix,superscriptaddress,longbibliography,  nofootinbib]{revtex4-1}
\RequirePackage[sort&compress]{natbib}

\usepackage{color}
\usepackage{appendix}

\usepackage{amssymb,amsfonts,amsmath}
\usepackage{bm}
\usepackage[pdftex]{graphicx}
\usepackage{xcolor}
\usepackage{comment}
\usepackage{float}
\usepackage[colorlinks = true,
linkcolor = blue,
            urlcolor  = blue,
            citecolor = blue,
            anchorcolor = blue]{hyperref}

\usepackage{hyperref}% add hypertext capabilities
\usepackage{dcolumn}% Align table columns on decimal point
\usepackage{bm}% bold math

\usepackage[caption=false]{subfig}
\usepackage{hyperref}
\usepackage{xr}
\makeatletter
\newcommand*{\addFileDependency}[1]{
  \typeout{(#1)}
  \@addtofilelist{#1}
  \IfFileExists{#1}{}{\typeout{No file #1.}}
}
\makeatother

\renewcommand{\(}{\left(}
\renewcommand{\)}{\right)}

\newcommand{\tr}[1]{\text{Tr}\(#1\)}
\renewcommand{\(}{\left(}
\renewcommand{\)}{\right)}
\def\@cite#1#2{$^{\mbox{\scriptsize #1\if@tempswa , #2\fi}}$}
\definecolor{RoyalBlue}{HTML}{4169e1}
\definecolor{ForestGreen}{HTML}{228b22}
\definecolor{DarkGreen}{HTML}{006400}
\usepackage[normalem]{ulem} % added by Carlo!

\usepackage{setspace}
\usepackage{comment}
\usepackage{float}
\usepackage{todonotes}

\newcommand{\arsham}[1]{{\color{black} #1}}

\begin{document}
%%TC:ignore
%\quickwordcount{main}
%\quickcharcount{main}
%\detailtexcount{main}
%%TC:endignore

\preprint{APS/123-QED}
%%TC:ignore
\title{Statistical physics of complex information dynamics}% Force line breaks with \\

\author{Arsham Ghavasieh} 
\affiliation{Fondazione Bruno Kessler, Via Sommarive 18, 38123 Povo (TN), Italy}

\author{Carlo Nicolini}
\affiliation{Center for Neuroscience and Cognitive Systems, Istituto Italiano di Tecnologia, Rovereto (TN), Italy}

\author{Manlio De Domenico}
\email[Corresponding author:~]{mdedomenico@fbk.eu}%
\affiliation{Fondazione Bruno Kessler, Via Sommarive 18, 38123 Povo (TN), Italy}

\date{\today}% It is always \today, today,

\begin{abstract}
The constituents of a complex system exchange information to function properly. Their signalling dynamics often leads to the appearance of emergent phenomena, such as phase transitions and collective behaviors. While information exchange has been widely modeled by means of distinct spreading processes --- such as continuous-time diffusion, random walks, synchronization and consensus --- on top of complex networks, a unified and physically-grounded framework to study information dynamics and gain insights about the macroscopic effects of microscopic interactions, is still eluding us. In this article, we present this framework in terms of a statistical field theory of information dynamics, unifying a range of dynamical processes governing the evolution of information on top of static or time varying structures. We show that information operators form a meaningful statistical ensemble and their superposition defines a density matrix that can be used for the analysis of complex dynamics. As a direct application, we show that the von Neumann entropy of the ensemble can be a measure of the functional diversity of complex systems, defined in terms of the functional differentiation of higher-order interactions among their components. Our results suggest that modularity and hierarchy, two key features of empirical complex systems --- from the human brain to social and urban networks --- play a key role to guarantee functional diversity and, consequently, are favored.
\end{abstract}

\maketitle
%\tableofcontents
%%TC:endignore
\section{Introduction}
Complex systems - i.e., large collections of alike agents exhibiting non-trivial collective behavior --- are ubiquitous in nature and are characterized by emergent phenomena, from flocks of birds~\protect\cite{bialek2012} and swarms \protect\cite{swarms2017} to large-scale electrical activity in the human brain~\protect\cite{amari1974, rolls2010noisy}. Despite the increasing convincing evidence disfavoring the traditional reductionist paradigm of studying the system's components in isolation, a fundamental understanding of the emergent phenomena due to the interplay between the underlying structure and the dynamics on the top of it still eludes us.
Therefore, a general framework to model and characterize information dynamics is needed to explain their rich collective dynamics and common emergent properties, such as resilience~\protect\cite{gao2016universal}, robustness~\protect\cite{cohen2010complex} and functional diversity ~\protect\cite{Luppi2019}.
On the one hand, it is known that non-trivial interactions among the constituents --- e.g., in terms of long-range interactions reducing topological distances~\protect\cite{Watts1998} or heterogeneity ~\protect\cite{Barabasi1999} --- are responsible for the macroscopic behavior of complex networks. On the other hand, even if the type of interactions differs from one system to another --- e.g. neurons use electrochemical signals in the human brain~\protect\cite{park2013structural}, while individuals send and receive text messages on the social media~\protect\cite{Watts2002,Stella2018}---, they can all be understood in terms of information exchange among system's units. 

The flow of information between components is restricted by the underlying structure in a way that the neighboring components have higher chance for exchanging information, while the possibility of communication with the distant components depends on topological factors and, simultaneously, on the way those structures interact with the dynamics to exhibit distinct patterns of information flow~\protect\cite{Harush2017}. Thus, to have a comprehensive picture of complex information dynamics, it is essential to consider the coupling between the structure and processes such as diffusion~\protect\cite{dedomenico2016physics,dedomenico2017diffusion}, random walks~\protect\cite{masuda2017random}, synchronization~\protect\cite{arenas2008synchronization} and consensus. In spite of the existing studies~\protect\cite{prldiffusion,Rosvall2008,Newman2005}, a general framework which unifies a range of possible processes and sheds light on the interplay between structure and dynamics and the large scale physics of complex phenomena still eludes us. Importantly, considering our limitations in tracking the activity of the large number of elements in real world systems, it is essential that the desired framework, rather, captures the macroscopic aspects of complex information dynamics, like the role of structure in hindering or facilitating the information flow~\protect\cite{prr2020} and the richness of information dynamics~\protect\cite{Yamamoto2018}.

In this article, we propose a field theory for complex information dynamics to unify a range of dynamical processes governing the evolution of information on top of static or time-varying structures. We introduce "information streams" as a set of operators that determine the directions of information flow. We show that they provide a meaningful statistical ensemble to construct the statistical physics of complex information dynamics. The framework unifies our knowledge of structure and dynamics and opens the doors for a wide range of future studies, from critical phenomena to robustness and functional diversity of systems, with applications in a variety of disciplines, including physics, neuroscience, systems biology and social sciences. As a direct application, we show that the mixedness of the ensemble, quantified by its Von Neumann entropy, can be understood as a measure of functional diversity, defined as the differentiation of interactions among the components.

\section{Complex information dynamics}
Usually, systems are modeled \arsham{as} networks where the elements and the connections between them are respectively pictured as nodes and links. Assume nodes are identified by canonical vectors $\langle i| (i= 1,2,...N$) and the connections are encoded in a time-varying operator $\hat{W}(t)$, which in the space of nodes --- defined by the canonical vectors $\langle i|$ --- represents the weighted and directed adjacency matrix --- i.e., the weight of the link from $i$-th to $j$-th node is given by $\langle i| \hat{W}(t) |j\rangle = W_{ij}(t)$.

We introduce the information field $\langle \phi(t)|$ written in the vectorial notation for convenience, in a way that the amount of field on top of $i$-th node at time $t$ reads $\langle{\phi}(t)|i\rangle$. The evolution of the information field in the most general form follows $\partial_{t}\langle\phi(t)|=F(t,\hat{W}(t),\langle\phi(t)|)$, which after linearization becomes
\begin{equation}\label{eq:master_linearized}
\partial_{t}\langle\phi(t)|=\langle\phi(t)| F(t,\hat{W}(t)).
\end{equation}

The solution of Eq.~\ref{eq:master_linearized} follows the form $\langle\phi(t)|=\langle\phi(0)| \hat{S}(t,\hat{W}(t))$, with the propagator $\hat{S}(t,\hat{W}(t))$ that can be written as $\hat{S}(t)$ for brevity. Depending on the choice, $F(t,\hat{W}(t))$ can describe a range of possible dynamics of the information field, such as continuous-time diffusion, random walks, consensus and synchronization~\protect\cite{PhysRevLett.110.028701,masuda2017random,dedomenico2017diffusion,arenas2008synchronization,PhysRevLett.110.028701} on top of static and time varying structures. 

To picture the information flow among the nodes, we consider the initial state $\langle \phi(0)|=\phi_{0}\langle  i | $, where $\phi_{0}$ is constant and represents the initial value of the field. Accordingly, the information flow from $i$-th node is $\langle\phi(t)|=\phi_{0} \langle i|  \hat{S}(t)$, and the information received by $j$-th node from $i$-th follows $\phi_{0}\langle i|\hat{S}(t)|j\rangle$. 

Often, the initial conditions of a specific problem are unknown: hence, we prefer a statistical approach based on a probabilistic initial condition $\langle \phi(0)|=\sum\limits_{i=1}^{N} p_{i} \phi_{0}  \langle i|$, where $p_{i}$ is the probability that $i$-th node is the information field seed.
Accordingly, the expected information flow from $i$-th node becomes $p_{i} \phi_{0} \langle i| \hat{S}(t)$ and the expected information received by $j$-th node from $i$-th node follows $p_{i} \phi_{0} \langle i| \hat{S}(t)|j\rangle$. 

Assuming the maximum lack of knowledge about the initial conditions, we assign a uniform probability distribution $p_{i}=1/N$. Consequently, it is straightforward to find the expected information flow from $i$-th node
\begin{eqnarray}\label{eq:flow}
\frac{\phi_{0}}{N} \langle i| \hat{S}(t),
\end{eqnarray}
and calculate the expected information received by $j$-th node from $i$-th
\begin{eqnarray}\label{eq:exchange}
\frac{\phi_{0}}{N}  \langle i| \hat{S}(t)|j\rangle.
\end{eqnarray}

\section{Information streams}
A diagonalizable propagator can be written as $\hat{S}(t)= \sum\limits_{\ell=1}^{N} s_{\ell}(t)\hat{\sigma}^{(\ell)}(t)$, where $s_{\ell}(t)$ is the $\ell$-th eigenvalue of propagator, and $\hat{\sigma}^{(\ell)}(t)$ is the outer product of its $\ell$-th right and left eigenvectors. Consequently, the expected information exchange between pairs can be written as summation of $N$ different fluxes $\sum\limits_{\ell=1}^{N}\frac{\phi_{0}}{N}s_{\ell}(t)\langle i| \hat{\sigma}^{(\ell)}(t)|j\rangle$. The fluxes are directed between the nodes by a set of operators $\{\hat{\sigma}^{(\ell)}(t)\}$, acting as information streams. Each information stream is multiplied by a corresponding coefficient $\frac{\phi_{0}}{N}s_{1}(t),\frac{\phi_{0}}{N}s_{2}(t),...\frac{\phi_{0}}{N}s_{N}(t)$, which can be interpreted as the stream's size. Depending on size, each stream is considered inactive ($\frac{\phi_{0}}{N}s_{\ell}(t)=0$) or active ($\frac{\phi_{0}}{N}s_{\ell}(t)>0$). In the following we outline the role of information field in activating the streams and therefore, generating the flow.

\begin{figure}
%\centering
\includegraphics[width=0.5\textwidth]{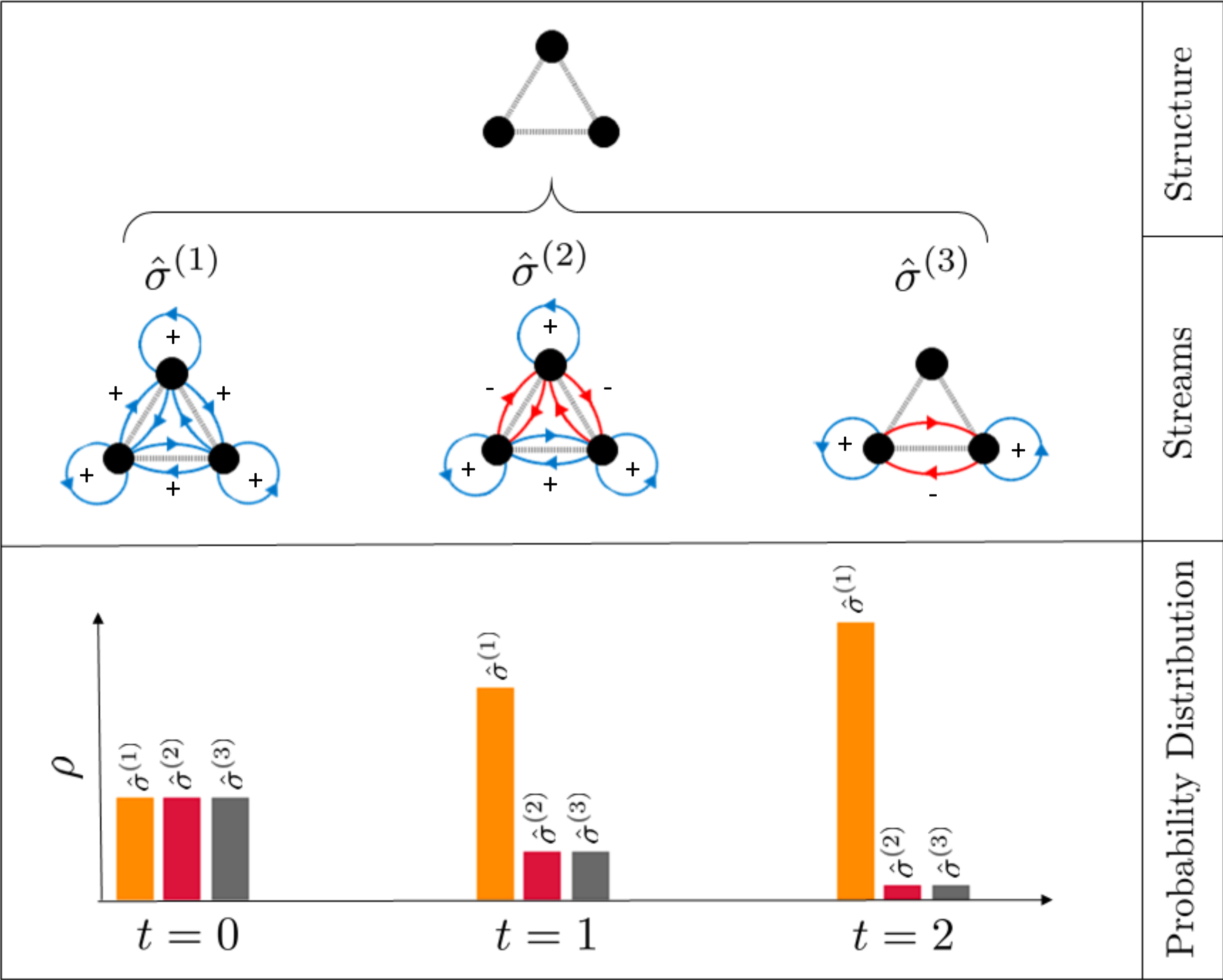}
\caption{\label{fig:ensemble}\textbf{Ensemble of information streams} A simple system of 3 fully connected constituents. The dynamical process is chosen to be random walk dynamics (see appendix.~\ref{app:RW}). The figure illustrates the information streams and their corresponding activation probabilities changing over time. 
}
\end{figure}

In the space of nodes, information streams can be represented by diagrams (see Fig~\ref{fig:ensemble}), where blue ($+$) and red ($-$) arrows, respectively, represent positive and negative fluxes guided by the streams --- e.g. $\langle i| \hat{\sigma}^{(\ell)}(t)|j\rangle$, depicted by an arrow from $i$-th to $j$-th node in the diagram of $\ell$-th stream. It is worth mentioning that in some cases, depending on the dynamical process, the field can not take negative values. Therefore, negative flux might be interpreted as reverse flux, forcing back the field carried by other information streams to the initiator node. Moreover, self-loops might trap a part of information field on top of the initiator node, e.g. $\langle i| \hat{\sigma}^{(\ell)}(t)|i\rangle$ depicted by a self-loop from $i$-th node to itself, in the diagram of $\ell$-th stream. 

\section{Activation of information streams}
For special case of random walk dynamics, the expected trapped field has been previously shown to be proportional to the average return probability of random walks~\protect\cite{prr2020}. Regardless of the type of dynamics, self-loops existing in information streams are directly responsible for trapping the field. Interestingly, it can be demonstrated (see appendix.~\ref{app:TF}) that the amount of field each information stream traps is equal to the size of that stream $\frac{\phi_{0}}{N}s_{\ell}(t)$, and consequently, the overall expected trapped field can be obtained from the summation of stream sizes $\sum\limits_{\ell=1}^{N}\frac{\phi_{0}}{N}s_{\ell}(t)=\frac{\phi_{0}}{N}Z(t)$, where $Z(t)=\sum\limits_{\ell=1}^{N}s_{\ell}(t)=\tr{\hat{S}(t)}$.

From another perspective, since the expected trapped field regulates the size of information streams, it can be considered responsible for activation of the streams, i.e. generation of flow. This understanding suggests that the dynamics of information field might be reducible to the dynamics of the trapped field. Consequently, using a proper superposition of information streams, $\hat{\rho}(t)=\sum\limits_{\ell=1}^{N} \rho_{\ell}(t) \hat{\sigma}^{(\ell)}(t)=\frac{\hat{S}(t)}{\tr{\hat{S}(t)}}$, where the $\ell$-th stream is weighted by its fractional share from the trapped field, $\rho_{\ell}(t)=\frac{\frac{\phi_{0}}{N}s_{\ell}(t)}{\frac{\phi_{0}}{N}Z(t)}$, we obtain another description of the expected information flow (Eq.~\ref{eq:flow}), reflecting the property
\begin{eqnarray}
\frac{\phi_{0}}{N} \langle i| \hat{S}(t) &=& \frac{\phi_{0}}{N}Z(t) \langle i| \hat{\rho}(t).
\end{eqnarray}

\section{Ensemble of information streams}
These weights, despite being fractions of the trapped field, admit a probabilistic interpretation. Assume the field is discretized into a large number of infinitesimal quanta carrying value $h$ which, depending on the nature of information field, can be bits of information, small packets of energy, infinitesimal volumes of fluid, etc. Consequently, the number $n(t)$ of quanta participating in activation of information streams is given by $n(t)h=\frac{\phi_{0}}{N}Z(t)$ and, similarly, the number $n^{(\ell)}(t)$ of quanta participating in activation of $\ell$-th information stream is $n^{(\ell)}(t)h=\frac{\phi_{0}}{N}s_{\ell}(t)$. Therefore, the probability that one quantum participates in activation of $\ell$-th stream is given by $\frac{n^{(\ell)}(t)}{n(t)}=\frac{\frac{\phi_{0}}{N}s_{\ell}(t)}{\frac{\phi_{0}}{N}Z(t)}=\rho_{\ell}(t)$. Note that this probabilistic interpretation is valid only if the eigenvalues of the propagator are real. Finally, each quantum of the trapped field generates a unit flow
\begin{eqnarray}\label{eq:unit_flow}
\frac{1}{n(t)} \frac{\phi_{0}}{N}Z(t) \langle i| \hat{\rho}(t)= h \langle i| \hat{\rho}(t),
\end{eqnarray}
by activating one of the information streams according to their activation probabilities. \arsham{The unit flow is the smallest element used to describe} functional interactions in the system. The expected information flow (Eq.~\ref{eq:flow}) is obtained from summation of all the unit flows. 

Hence, we can completely describe the information dynamics using a number of unit flows, each activating one of the information streams $\{\hat{\sigma}^{(\ell)}(t)\}$. Considering the probabilistic nature of this process, the streams shape a statistical ensemble encoding all possible fluxes among components and their probabilities. Furthermore, the operator $\hat{\rho}(t)$ --- as the superposition of information streams weighted by their activation probabilities --- is reminiscent of the density matrix used in quantum statistical physics in terms of the superposition of quantum states.

\begin{figure*}[t]
    \centering
    \includegraphics[width=.8\textwidth]{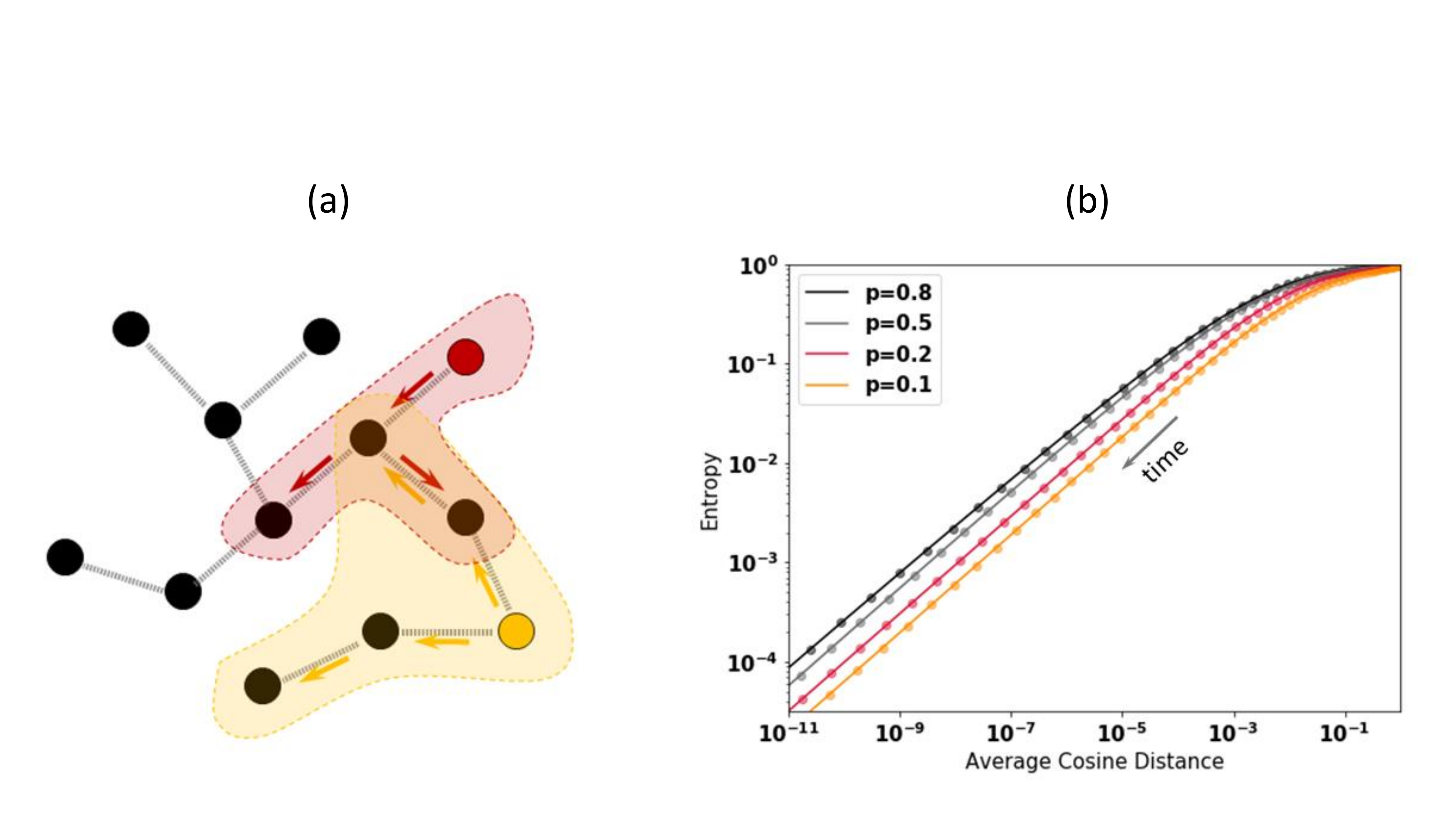}
    \caption{\label{fig:diversity}\textbf{Functional diversity.} (a) Overlap of the flow initiated from two nodes, shown in red (dark gray) and yellow (light gray). In case of full overlap, both nodes play highly similar roles, as senders of information. (b) Four Erdos-Renyi networks with different connectivity probabilities are considered to show the relationship between Von Neumann entropy, rescaled by its upper bound $\log N$, and the overlap of information flow initiated from different nodes, encoded by average cosine dissimilarity (see text for details). Lines correspond to networks with connectivity probabilities of 0.1, 0.2, 0.5 and 0.8, respectively from bottom to the top of the plot. }
\end{figure*}

\section{Von Neumann Entropy}
The unit flow, from $i$-th to $j$-th node, can be expanded in terms of all available fluxes multiplied by their activation probabilities, as $h\langle i| \hat{\rho}(t)|j\rangle=h\sum\limits_{\ell=1}^{N} \rho_{\ell}(t)  \langle i| \hat{\sigma}^{(\ell)}(t)|j\rangle $. The probability distribution plays a key role in the information flow between the nodes. For instance, a flat distribution $\rho_{\ell}(t)=1/N$ allows for activation of all information streams, indicating high diversity of available fluxes and information dynamics. The mixedness of information streams, inspired by quantum statistical mechanics, can be quantified by the Von Neumann entropy of the ensemble
\begin{eqnarray}\label{eq:entropy}
\mathcal{S}(t)= -\tr{ \hat{\rho}(t)\log{ \hat{\rho}(t)}}.
\end{eqnarray}{}

When the information dynamics is completely describable by one information stream, we obtain zero entropy. In contrast, as information dynamics becomes rich and diverse, the number of information streams required to capture it grows, and consequently, entropy takes higher values. The diversity of information dynamics is fundamental for a number of important properties of complex systems that we discuss in the next sections.

\arsham{Interestingly, when the dynamical process governing the evolution of information field is the continuous diffusion, the Von Neumann entropy of the ensemble coincides with the spectral entropy~\protect\cite{dedomenico2016spectral} proposed to analyze a variety of empirical complex networks from an information-theoretic perspective. Our field theory sheds light on the physical meaning of such methods and might even lead to deeper understanding of their results. For instance, spectral entropy has been successfully used to analyze 45 virus-human protein-protein interactomes and uncover non-trivial clusters of viruses~\protect\cite{ghavasieh2020multiscale}. Noticeably, in the mentioned study, the number of viral components was negligible compared to the number of targeted human proteins, leading to the failure of commonly adopted structural metrics in detecting the distinguishing features of different viruses. Conversely, the disruption of information dynamics among human proteins has been clearly reflected in mixedness of the streams, captured by the Von Neumann entropy and used to cluster together viruses based on their effects.}

\section{Functional diversity}
Although a system's units are and behave similar when considered in isolation, they exhibit a variety of functional roles with the system: e.g., neurons belonging to two different functional areas of the human brain might be involved in distinct cognitive activities, as a consequence of their differences in handling information, as sender, receiver and processor~\protect\cite{kotter2003}. In this framework, the information flow initiated from a node is encoded by a vector given by Eq.~\ref{eq:flow}. In addition to being a sender, each node receives information from the others encoded by another vector, which for $i$-th node reads $\frac{\phi_{0}}{N} \hat{S}(t)|i\rangle$. Statistical analysis of sent and received information vectors, corresponding to the nodes, can be used to asses how similar or functionally diverse they are. In the following, we show that the diversity of functional roles among the nodes is reflected in the overall functional diversity quantified by Von Neumann entropy.

In the following, we analytically  show that Von Neumann entropy is sensitive to the cosine dissimilarity of sent and received information vectors corresponding to each node, interpreted as the dissimilarity of the two roles each node plays as sender and receiver.

\arsham{Remind that the outgoing unit flow from $i$-th constituent reads $h\langle i| \hat{\rho}(t)$ (See Eq.~\ref{eq:unit_flow}). Therefore, the incoming unit flow to the same node can be obtained as $h\hat{\rho}(t)|i\rangle $. Evidently, the inner product of the two vectors is given by $ h\langle i| \hat{\rho}(t) h \hat{\rho}(t)|i\rangle = h^{2} \langle i| \hat{\rho}^{2}(t) |i\rangle  $, depending on the diagonal elements of the second power of the density matrix. 

When the Von Neumann entropy is approximated as linear entropy, the diagonal elements of $\rho^{2}(t)$ become determinant  
\begin{eqnarray}
\mathcal{S}(t) &\approx&  1 - \tr{ \hat{\rho}^{2}(t)}.
\end{eqnarray}{}

Thus, one can conclude that the second term of the linear entropy is inversely proportional to the summation of inner product of outgoing and incoming unit flow,  $\tr{ \hat{\rho}^{2}(t)}=\frac{1}{h^{2}}\sum\limits_{i=1}^{N}h^{2}\langle i|\hat{\rho}^{2}(t)|i\rangle$. The argument of the summation, as an inner product, depend on magnitude of each flow and their cosine similarity. The former indicates the magnitude of interactions while the latter encodes the symmetric role of each constituent as sender and receiver of information. Consequently, entropy favors lower cosine similarity, or in other words the asymmetric role of constituents as sender and receiver.

In the same spirit, nodes can play different roles with respect to one another. In fact, cosine dissimilarity of the sent information vectors corresponding to each pair of nodes can be a proxy of their functional difference, as senders. The Von Neumann entropy is expected to be sensitive to this type of functional diversity, as well.

To verify it, we consider three types of dynamical processes governing the evolution of information field, including random walk dynamics, the dynamics of coupled oscillators close to meta-stable synchronized state and the consensus dynamics, which are all governed by normalized Laplacian, following Eq.~\ref{eq:master_linearized}, and, consequently, a similar propagator, stream size distribution and Von Neumann entropy (see appendix.~\ref{app:RW} for details).

In random walk dynamics, the information field encodes the density of random walkers on top of nodes and the expected information exchange between two nodes is the number of random walkers traveling from first node to the other one. In consensus dynamics, it is the number of individuals reaching consensus on top of the nodes and the information exchange is the impact of one node on another towards their consensus. Similarly, the described dynamics remains valid for coupled oscillators in the meta-stable synchronized state, where the information field encodes the phase deviation of the oscillators from synchronized state, and how the deviation propagates into the network is the information flow. For all the above cases, the density matrix reads $\hat{\rho}(t)=e^{-t \hat{L}^{*}}/Z(t)$, where $ \hat{L}^{*}$ is the normalized Laplacian operator or its transpose.

Following Eq.~\ref{eq:unit_flow}, we indicate the outgoing unit flow vectors corresponding to $i$-th and $j$-th nodes by $\langle \phi^{(i)}(t)| = \frac{h}{Z(t)}\langle i|e^{-t\hat{L}^{*}}$ and  $\langle \phi^{(j)}(t)| = \frac{h}{Z(t)}\langle j|e^{-t\hat{L}^{*}}$ and, consequently, their cosine dissimilarity reads
\begin{eqnarray}
d_{ij}(t)=1-\frac{  \langle \phi^{(i)}(t)| \phi^{(j)}(t)\rangle  }{  \sqrt{ \langle \phi^{(i)}(t)| \phi^{(i)}(t)\rangle\langle \phi^{(j)}(t)| \phi^{(j)}(t)\rangle}  }.
\end{eqnarray}{}
The average cosine dissimilarity given by $\bar{d}(t)=\frac{1}{N^{2}}\sum\limits_{i,j=1}^{N}d_{ij}(t)$ can be used as a proxy of functional diversity of nodes, as senders.} Here, we \arsham{numerically} show that the average cosine dissimilarity is proportional to the Von Neumann entropy of the statistical ensemble $\rho(t)$ (Fig.\ref{fig:diversity}). \arsham{Although unit flows are used in this section's analysis, the results remains valid for the full information flow vectors given by Eq.~\ref{eq:flow}, as these vectors are equal to unit flow vectors multiplied by the amount of trapped field and multipliers do not affect the cosine dissimilarity. }

Remarkably, low entropy indicates high overlap between the flow vectors initiated from nodes, implying that they play rather similar roles as senders, on average. In such an interconnected system, random failures or targeted attacks can cause negligible damage, as the remaining nodes are, likely, able to replace and send information to the group of nodes accessible by the removed ones. 

\begin{figure*}
\centering
\includegraphics[width=.91\textwidth]{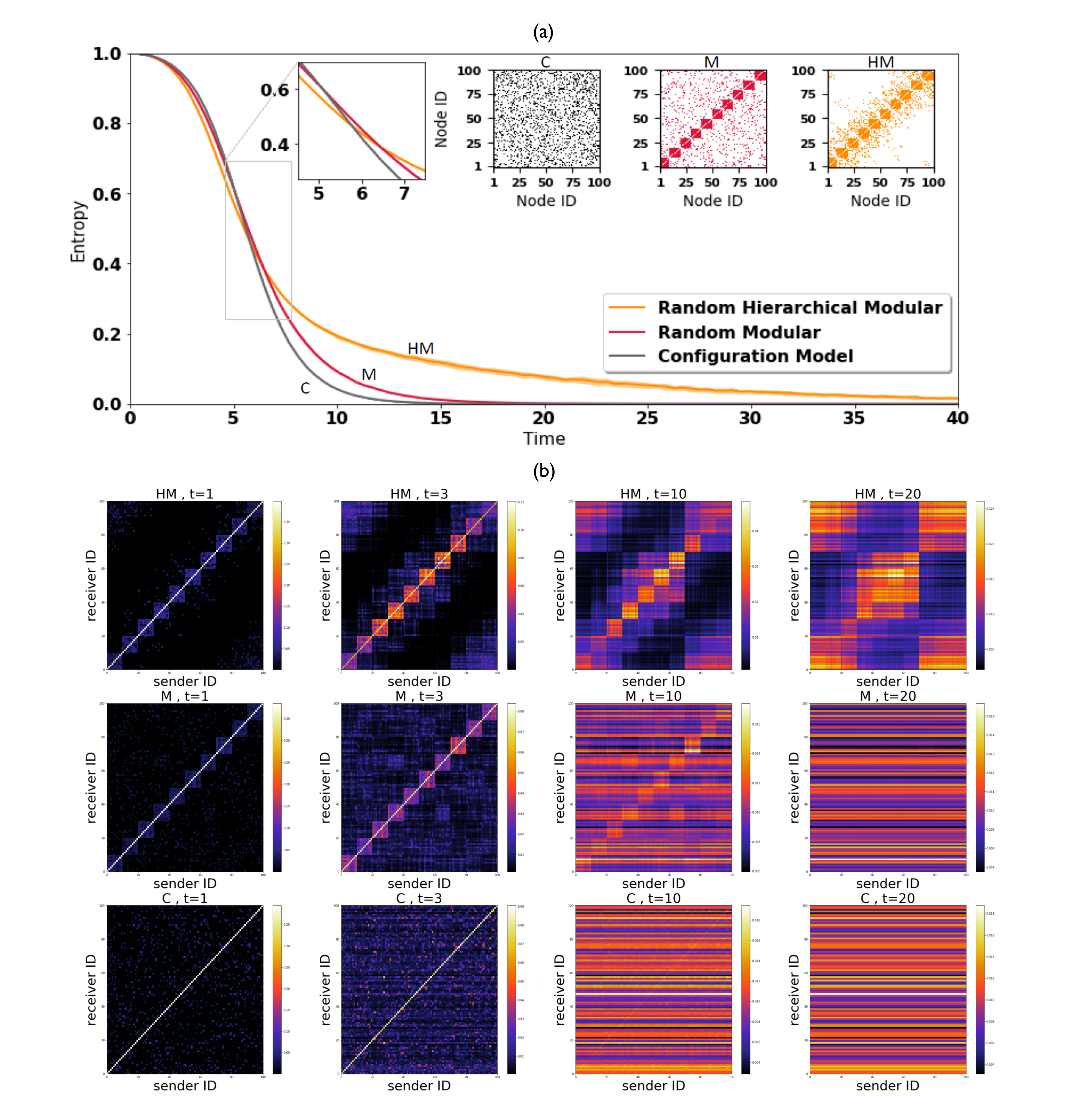}
\caption{\label{fig:modular}\textbf{Modularity and hierarchy} (a) The entropy, rescaled by its upper bound $\log{N}$, of hierarchical modular topology(HM), modular topology(M), and their configuration model(C), which randomizes the links while keeping the degree distribution, is compared. The adjacency matrix of a realization of each topology is shown in top panels. For each topology, 100 realizations have been considered, and their mean entropy is plotted by solid lines with shades, representing the fluctuations around the mean. Our result shows that random topology exhibits slightly higher entropy at small time scales, while at larger time scales, the modular and hierarchical modular topology exceed it. Interestingly, hierarchical modular topology with significantly slow decay, preserves its functional diversity across a considerable range of time scales. \arsham{(b) Node-node communications are plotted as heatmaps for one realization of each considered topology at four different propagation times $t=1,3,10,20$. Each value, represented as a pixel of a heatmap, corresponds to information exchange between two nodes ---e.g. $\langle i | \hat{S}(t)|j\rangle$ for nodes $i$-th and $j$-th. In hierarchical modular networks, long range communications, characterized by large propagation times, are meaningful as nodes keep their functional diversity. In contrast, when connections are randomly distributed, the field quickly reaches its final frozen state and the nodes can not be differentiated as senders.  }   }
\end{figure*}

\section{Hierarchy and modularity}

\arsham{In the last section, we extensively explored the relationship between the diversity of information dynamics, given by the mixedness of ensemble, and the diversity of the nodes, as senders and receivers of information at microscopic level. In this section we discuss the effect of a number of large-scale network features on the Von Neumann entropy, showing that the presence of these topological complexity can keep information streams alive and mixed even at large propagation time-scales, allowing units to communicate and take diverse functional roles. In fact, as we explain in the following, these two features impose restrictions on the information dynamics, by separating groups of units from each other and controlling the directions of flow between them, leading to a boost in functional diversity.}

Hierarchical and modular structures have been observed in a multitude of complex systems~\protect\cite{barabasi2003}, from neuroscience~\protect\cite{anderson2013describing}, to ecology~\protect\cite{hulot2000functional} and social sciences~\protect\cite{Watts2002}. These features are important for many vital properties of such systems --- e.g. in the human brain, it has been shown that they play a key role in expanding the range of critical neural dynamics~\protect\cite{Hilgetag2014}, diversity of dynamical patterns~\protect\cite{dynamicalpatterns}, robustness and adaptation~\protect\cite{Meunier2010}. Our results, shown in  Fig.~\ref{fig:modular}, unravel two important characteristics of systems exhibiting this type of structure. First, they exhibit rich and diverse information dynamics, at unexpectedly large temporal scales. While the size of information streams decay quickly in random and random modular networks, hierarchy keeps the streams active and the equilibrium distant to reach. Second, information flow vectors originated from the nodes are highly diverse as a consequence of modularity and hierarchy. In random networks, each node can exchange information with any other node according to a given probability. Whereas, in random modular structures information exchange occurs, mostly, inside the modules and therefore, the flow vectors initiated by nodes in different modules have low or no overlap. In addition to this constraint due to modularity, the information flow between the modules itself is influenced by the hierarchy, as the top modules have higher access to lower modules. Therefore, in hierarchical modular systems, the roles of nodes as senders or receivers of information vary, from node to node and from module to module which is, as expected, reflected in the Von Neumann entropy\arsham{(See Fig.~\ref{fig:modular} panel (b))}. 

In hierarchical modular systems, even at large time scales, the dynamics remains rich, diverse and distant from the frozen state --- note that, for the dynamics considered here, this is reached in the limit for infinite time, which exhibits the same physical properties of a state at zero absolute temperature ---, keeping the random walkers, the state of consensus and the phase of oscillators far from their final distribution. Additionally, the distribution of field, even at large time scales, \arsham{is highly dependent on its initial conditions}, determined by the node that initiates the dynamics. 

\section{Conclusion}

\arsham{Despite the deep knowledge of structure gained during last decades, and the important steps taken towards understanding the coupling between structure and dynamics~\protect\cite{prldiffusion,prr2020,Harush2017}, there has been a lack for a unifying framework to describe information dynamics within interconnected systems and the complex emergent phenomena at macro-scale.} 

The aim of this work is to provide a theoretical framework to fill the gap, providing a ground to better understand, through statistical physics, phenomena observed in complexity science. To achieve this purpose, we did not rely on the blind application of existing mathematical machinery of statistical physics: instead, we have developed a consistent statistical field theory where  stream operators allow one to understand information flow and its trapping in the structure, while the corresponding probabilities define a novel statistical ensemble.

In fact, our framework allows one to better understand the physical meaning of and derive a physical ground for the density matrices --- based on continuous diffusion~\protect\cite{dedomenico2016spectral} and random walk dynamics~\protect\cite{prr2020} --- recently used to analyze complex networks from an information-theoretic point of view. Remarkably, it also allows for generalizations to other classes of dynamics which, at first order, can be approximated by linear master equations governed by any function of the adjacency matrix, \arsham{and used to analyze a wide variety of complex interconnected systems, when unit-unit interactions are important. }

\arsham{Of course, complex information dynamics is in the realm of non-equilibrium physics. Yet, we have shown that it can be fully characterized in terms of Gibbsian-like density matrices, allowing for direct implementation of a wide range of methods and techniques developed in quantum statistical physics. Furthermore, following the proposed framework, the state of a system describes short to long range interactions among its units, depending on the choice of the propagation time $t$. In turn, tuning the propagation time allows for multiscale analysis of node-node communications. More importantly, the proposed theory paves the way for better-understanding the non-trivial higher order interactions, from a physical perspective. 

As a direct application, we demonstrated the relationship between the mixedness of ensemble and the functional diversity of nodes, as senders and receivers of information. This gives insights about how the agents of complex systems which are often similar in isolation, can play distinct functional roles in the system --- e.g. neurons belonging to two different functional areas of the brain involving in different tasks. Furthermore, our findings provide ground for better-understanding the recently proposed definition of node-network entanglement~\protect\cite{Ghavasieh2020UnravelingTE}, quantified as the alteration of Von Neumann entropy due to detachment of nodes. It is worth mentioning that entanglement has been used as a multiscale centrality measure, laying out a novel effective attack strategy to disintegrate complex networks. One can exploit the simple fact that  a node has higher entanglement if its removal affects the system's functional diversity more severely. Its effects can be analyzed by calculating the difference, in terms of the Von Neumann entropy, between the original system and the perturbed system consisting of the detached node with its edges forming a star graph plus the remainder of the original network. This knowledge encourages further studies regarding the effect of network contraction on the node-node communications (See Eq.~\ref{eq:exchange}) and functional diversity (See Eq.~\ref{eq:entropy}) at different temporal scales, to push the boundaries of robustness studies beyond the traditional analysis. Also, our theoretical findings provide physical insights about other already explored applications of Gibbsian-like density matrices, from network clustering to transport phenomena~\protect\cite{prr2020}. Since information dynamics play a crucial role in a wide range of problems related to complex systems, our field theory provides a useful ground for a range of future theoretical developments.
}

Finally, it is worth noting that when the network is quantum and its dynamics is governed by a quantum Hamiltonian $\hat{H}$, our stream operators become the outer product of left and right eigenvectors of the Hamiltonian and our density matrix becomes physically consistent with the density matrix governing quantum statistical mechanics at equilibrium after Wick rotation. This result opens the door to additional interesting future perspectives.

%%TC:ignore
\begin{acknowledgements} 
C.N has received funding from the European Union’s Horizon 2020 Research and Innovation Program under grant agreement 668863, SyBil-AA.
\end{acknowledgements}
%%TC:endignore

\appendix

\section{Stream sizes and the trapped field}\label{app:TF} 

The expected trapped field, at a given time, can be written as summation of expected trapped field on top of all the nodes
\begin{eqnarray}
\frac{\phi_{0}}{N}\sum\limits_{i=1}^{N} \langle i| \hat{S}(t) |i \rangle =\frac{\phi_{0}}{N} \tr{S(t)}, \nonumber
\end{eqnarray}{}
which, after eigen-decomposition of the propagator becomes
\begin{eqnarray}
\frac{\phi_{0}}{N}  \tr{S(t)}  =  \frac{\phi_{0}}{N} \sum\limits_{\ell=1}^{N} s_{\ell}(t) \tr{\hat{\sigma}^{(\ell)}(t)}, \nonumber
\end{eqnarray}{}
where $\frac{\phi_{0}}{N} s_{\ell}(t) \tr{\hat{\sigma}^{(\ell)}(t)}$ is the expected amount of field the $\ell$-th stream traps at time $t$. Since $\hat{\sigma}^{(\ell)}(t)$ is the outer product of the $\ell$-th left and right eigenvectors of the propagator, its trace is one. Therefore, the expected amount of field trapped by the $\ell$-th stream follows 
\begin{eqnarray}
\frac{\phi_{0}}{N} s_{\ell}(t) \tr{\hat{\sigma}^{(\ell)}(t)} = \frac{\phi_{0}}{N} s_{\ell}(t), \nonumber
\end{eqnarray}{}
which is equal to the size of the stream.

\section{Dynamical processes}\label{app:RW} 

\arsham{In this section, we derive the density matrix for three examples of dynamical processes including random walkers, consensus formation and dynamical coupled oscillators. Although these dynamics are of distinct nature, they are all governed by the normalized Laplacian.~\protect\cite{dedomenico2017diffusion}}

1. A random walker jumps from node $i$ to node $j$ with probability $\langle i| \hat{T}|j\rangle = T_{ij}=\frac{W_{ij}}{k_{i}}$, where $\hat{T}$ is the transition operator of the random walk, and $\hat{W}$ is the adjacency operator and $k_{i}=\sum\limits_{j=1}^{N} W_{ij}$ is degree (strength) of node $i$. Having the transition matrix, we can find the normalized Laplacian given by $\hat{L}^{*}=\hat{I}-\hat{T}$. 
If the $i$-th components of the information field, $\langle \phi(t)|$, indicates the probability to find the random walker on node $i$ at time $t$, its evolution is governed by the master equation
\begin{eqnarray}
\langle \phi(t+1)|= \langle \phi(t)| \hat{T}
\end{eqnarray}

which, in the continuous-time approximation reduces to 
\begin{eqnarray}
\partial_{t}\langle \phi(t)|=-\langle \phi(t)|\hat{L}^{*} 
\end{eqnarray}

with solution given by $\langle \phi(t)|=\langle \phi(0)|\hat{S}(t)$, where $\hat{S}(t)=e^{-t\hat{L}^{*}}$.

The propagator corresponding to the normalized Laplacian, $\hat{S}(t)=e^{-t\hat{L}^{*}}$ (For complete derivation in case of RW see~\protect\cite{RWnoah}), is used to construct the density matrix $\hat{\rho}(t)=\frac{e^{-t\hat{L}^{*}}}{\tr{e^{-t\hat{L}^{*}}}}$ based on our framework.

2. Network of coupled oscillators are generally governed by Kuromato dynamics, given by
\begin{eqnarray}
\partial_{t}\langle \phi(t)| i \rangle = \omega_{i}(t) + \sum\limits_{j=1}^{N} \sigma_{ij} W_{ij} \sin( \langle \phi(t)| j \rangle - \langle \phi(t)| i \rangle  ) \nonumber
\end{eqnarray}
where $\langle \phi(t)| i \rangle$ is the phase of $i$-th oscillator at time $t$, $\omega_{i}(t)$ is its natural frequency and the mixing rate can be considered to be $\sigma_{ij}(t) = K/k_{i} $ ($K$ and $k_{i}$ being the coupling constant and the degree of $i$-th node, respectively). Assuming the same natural frequency for all oscillators, all oscillators will have the same phase at the equilibrium, $\langle\phi(t)|i\rangle = \langle\phi(t)|j\rangle$. The phase of each oscillator might deviate from the synchronized state, and consequently, the deviation propagates into the system. Considering small deviations $\langle\phi(t)|i\rangle \approx \langle\phi(t)|j\rangle$ the above equation can be approximated using $ \sin(\langle \phi(t)| j \rangle - \langle \phi(t)| i \rangle )\approx \langle \langle \phi(t)| j \rangle - \phi(t)| i \rangle $. Under this regime, the Kuromato model reduces to 
\begin{eqnarray}
\partial_{t}\langle \phi(t)| i \rangle = \sum\limits_{j=1}^{N} K \frac{W_{ij}}{k_{i}}  [\langle \phi(t)| j \rangle - \langle \phi(t)| i \rangle ], 
\end{eqnarray}
which, taking $K=1$, in the vectorial form reads
\begin{eqnarray}
\partial_{t}\langle \phi(t)|=-\langle \phi(t)|(\hat{I} - \hat{T^{\dagger}}) = -\langle \phi(t)|\hat{L}^{*\dagger}.
\end{eqnarray}

3. The consensus dynamics describes the tendency of n integrator units towards reaching an agreement. We represent the state of consensus of node $i$-th as $\langle \phi(t)| i \rangle$ which evolves following diffusive dynamics. To model this type of dynamics, it is suggested to control the large effect of hubs on other nodes, for instance by rescaling the weights of link emanating from each node by their degree. This leads to a continuous-time DeGroot model which can mathematically described as 

\begin{eqnarray}
\partial_{t}\langle \phi(t)| i \rangle = \sum\limits_{j=1}^{N} \frac{W_{ij}}{k_{i}}  [\langle \phi(t)| j \rangle - \langle \phi(t)| i \rangle], 
\end{eqnarray}

that in the vectorial form reads 
\begin{eqnarray}
\partial_{t}\langle \phi(t)|=-\langle \phi(t)|(\hat{I} - \hat{T}) = -\langle \phi(t)|\hat{L}^{*\dagger},
\end{eqnarray}

\bibliographystyle{apsrev}
%%TC:ignore
\bibliography{biblio}
%%TC:endignore

\end{document}